\documentstyle[twocolumn,aps]{revtex}
\begin{document}

\begin{title} 
{RVB Revisited
}
\end{title}

\author{ Philip W. Anderson}

\address {Joseph Henry Laboratories of Physics\\
Princeton University, Princeton, NJ 08544}
\maketitle

\vskip.5truein

\abstract
We propose that early speculations on the nature of the normal
state of the $CuO_2$ planes in high $T_c$ superconductors can be
revised and brought into agreement with more modern theories,
simulations and experimental observations. The early speculations 
proposed an ``RVB'' state which is an incompressible liquid of singlet pairs in
which the magnetic excitations are spin 1/2 fermion-like objects
called spinons, the charge exhibiting a separate dynamics
(charge-spin separation.)  The revisions proposed are the
following: (1) At finite concentrations $x$ of holes 
there are two non-Fermi liquid, charge-spin separated states,
(I) a metal with a full Fermi surface (the ``Luttinger Liquid'')
and no spin gap, and below a crossover (II) a spin-gapped RVB
state of ``d'' symmetry with spinons at gap nodes (``nodons'').

(2) Since spinons have no phase---no charge---the gap parameter
in the second phase must be real, $d_{x^2-y^2}$, as proposed by
Affleck and Kotliar.

(3) Both are based on a large Fermi surface satisfying Luttinger's
theorem. 

(4) Phase I is a good metal in the $ab$ direction, phase II is a
poor metal. There appear to be two types of transition into the
superconducting state, depending on whether it takes place from
phase I or II. Whether there are two mechanisms causing
superconductivity remains unclear as also is the relationship
between the two.

In the first months after the discovery of high $T_c$
superconductivity the close connection of this
phenomenon with the one-band Hubbard model in two
dimensions was realized \cite{1,2}. An old speculation called ``resonating valence
bonds'', about a possible non-symmetry-breaking solution of
the antiferromagnetic Heisenberg model in two dimensions (to which the
Hubbard model reduces at half filling), was revived\cite{1}, and
used to motivate a scenario for explaining the high Tc
phenomenon.

Zou and Baskaran, along with others\cite{3}, showed that the RVB
state naturally led to the idea of charge-spin separation and to
the concept of the ``spinon'', a spin 1/2 object which can be
thought of as half of a spin wave or half of an electron (the idea of Fermionic spin 1/2 
excitations of an antiferromagnet dates back in fact to Landau.) The
one-dimensional Hubbard model has non-classical soliton-like
excitations carrying a spin of 1/2, and it was thought that they
could appear in 2D also. The ``RVB'' state is a ground
(``vacuum'') state made up of singlet pairs of electrons, and the
spinon soliton is thought of as a single spin moving freely
through the background gas of singlets, displacing the singlet
correlations but not the charge. In one D it can also be thought of as a
$\pi$ phase singularity in a spin density wave. The RVB state is
an incompressible spin liquid, in the sense that the one-electron 
compressibility (the density of states) vanishes, ${\partial \Sigma\over \partial k}\ {\rm
and}\ {\partial \Sigma\over \partial\omega}$ both diverge, and
$v_F$ and hence the true compressibility  
remains finite but $Z\equiv 0$.   

The most accurate characterization of the spinon has come from
studies of exactly soluble one-dimensional models, the Lieb-Wu
solution of the Hubbard model and the Haldane-Shastry solution of
the $1/r^2$ model, as well as from tomographic bosonization of
the 2D interacting electron gas. The spinon can be defined in
terms of non-Abelian bosonization of the spin fluctuations of a
Fermi sea at any interaction and for any density, and is
necessarily a semion in Haldane's sense rather than an anyon or
a true Fermion. For any case where renormalized perturbation
theory (Fermi Liquid Theory) remains valid, its velocity is
degenerate with that of the charge fluctuations and it is merely
half of a quasiparticle. It is also useful to think of it as
moving on a ``squeezed'' Heisenberg model from which the holes
have been removed; at the Fermi surface particles move 
non-diffractively and the spin degrees of freedom are like
those in the 1D model because removing holes does not alter the
spin ordering. 

While the spinon, at least in 2D, is rigorous only as a Fermi
surface excitation, visualization in terms of the RVB type of
state as a Gutzwiller-projected electron, as in Ref. 2, is a
useful guide to our thinking. Of course, a similar caveat about
the Fermi surface 
exists for the quasi-particle concept, for that matter. 

It soon became clear that the RVB idea in its original form was
not sophisticated enough to account for high Tc. While 
other authors, most notably Lee and co-workers\cite{4}, continued with
modified versions of the original scheme, a number of authors
beginning with R.B. Laughlin\cite{5} and including
Wilczek\cite{6} and others
have emphasized the ``flux phase'' version of RVB and its
relation to anyon theory, while I have retained the idea of
spinons and spin-charge separation, but focussed on the non-Fermi
Liquid metallic state\cite{7}, and on the effect of spin-charge
separation in modifying transport processes, and neglected the
``pairing'' aspect of the theory. In fact, I postulated that the
single 2D plane was not superconducting by itself\cite{7}

It was in connection with RVB theory that the idea of a d-wave
order parameter first arose, in work of Affleck and of
Kotliar,\cite{8}
since in a simple square lattice Heisenberg model ``$d$ wave'' is a
natural solution of the mean field equations using
nearest-neighbor antiferromagnetic exchange
$$\sum_{(ij)=nn} J_{ij}\ S_i\cdot S_j\eqno(1)$$
as a 4-spinon interaction, where
$$\vec S_i=\sum_{\sigma\sigma'} s^+_\sigma \ \vec S_{\sigma\sigma'}
\ s_{\sigma'}.\eqno(2)$$
The $s$'s are spinon operators, which we shall define later.

The basic derivation of (1) was put 
forward by Rice, adapted from my own derivation of
``superexchange'' in Mott insulators. A modern version to which
we will later refer was given by Zou. In the presence of a strong
repulsive interaction ``U'', the amplitude of doubly occuped
states is renormalized to zero by a projective canonical
transformation which is perturbative in $t/U$, where $t$ is the
hopping kinetic energy. This projective transformation cannot be
continued smoothly to a perturbation theory in positive powers
of the interaction $U$ because it is projective. Thus when an antiferromagnetic ``$J$'' is
postulated, Fermi liquid theory has per se been abandoned.
The signal for this is that the kinetic energy term ``$T$'' in
the $t-J$ Hamiltonian is projected onto a subspace with zero
double occupancy, so that the fluctuations are algebraically
distinct from conventional quasiparticles.

We observe that the spin fluctuations can be expressed in terms
of spinons, and that the exchange interaction affects only the
spinons directly, since the two sites ``$i$'' and ``$j$'' must
both be occupied in order for the superexchange effect to act. In
most theories of the cuprates, conventional exchange (which is
always ferromagnetic) is neglected. Thus we follow custom in
assuming that the true interaction is $S_i\cdot S_j-{n_in_j\over 4}$
and no interaction affects parallel spins, in our
first crude survey of the problem.

The original RVB hypothesis was that a good variational ground
state of (1) could be attained by the Ansatz
$$\psi_0=P^A_G \psi_{MF}\eqno(3)$$
where $P^A_G$ is the $n=1$ projector
$$P^A_G = \prod_i\ (n_{i\uparrow}+n_{i\downarrow}-2n_{i\uparrow}\ n_{i\downarrow})\eqno(4)$$
and $\psi_{MF}$ is a mean field BCS or Hartree-Fock solution of
the problem of N electrons on N sites. My original hypothesis was
a BCS solution with a symmetrical energy gap going to zero at a
Fermi surface, with 
kinetic energy,
$$\Delta_k=\gamma_k=J(cos\, k_x+cos\, k_y)\eqno(5)$$
which is called an ``extended $s$'' solution, and is equivalent,
as I showed (following Rice) to a half-filled tight-binding band of noninteracting
electrons, projected on $n_i=1$. 

Affleck soon pointed out that an equivalent solution
was the `$d$-wave'' with 

$$\Delta^d_k=J (cos\, k_x - cos \,k_y)\eqno(6)$$
and shortly thereafter Kotliar, as well as Zhang et al, suggested ``$s+id$'' where
$$\Delta_k= \gamma_k+i\Delta^d_k\eqno(7)$$
which is variationally better and has nodes of the one-spinon
energy rather
than a Fermi surface.  This in turn was equivalent,
essentially, to another choice, the so-called ``flux phase'', in
which the Hartree-Fock solution is for electrons with kinetic
energy $\epsilon_k=\gamma_k$ and a flux of $\Phi_{0/2}=$``$\pi$''
natural units through each plaquette of the lattice.

All of these ``RVB'' solutions of the Heisenberg Hamiltonian (1)
could be variationally improved and their energies
compared, but all turned out to be slightly inferior to the
N\'eel state with antiferromagnetic long range order, which also
turned out to be the actual physical state of the cuprate $CuO_2$
planes when doped to $Cu^{++}$, where (1) is an appropriate
Hamiltonian. Hsu showed that one could consider the N\'eel
state as a somewhat weak symmetry-breaking perturbation of the
flux phase, and understand its experimental high-energy spectrum in
terms of flux phase spinons, but otherwise the ``flux phase'' has
receded somewhat into the background as  a result of the fact that
it seems not to be easily generalized to the doped 2D case with
mobile holes, which is thought to be a good model for the
superconducting cuprates. The most straightforward such attempt
is to change the flux from $\pi$, giving a state which is not
time-reversal invariant, contrary to all experimental
tests.

As a result of this failure, the flux phase solution has tended
to be ignored, even though it has the nodal structure which is so
conspicuous in the observations, and which shows up fairly
clearly in numerical calculations on the 2D Hubbard
model\cite{9}
A purpose of the present paper is to suggest that the correct
generalization of the ``$s+id$'' or flux phase does not involve
time-reversal and that the apparent dependence upon the
introduction of ``i'' or a $\pi$ flux is illusory. This was 
indicated very early by constructing actual real space wave
functions for the flux phase, which turned out to be entirely
real. 

In the early work spinons were treated as full-fledged Fermions.
In order to satisfy the projective constraints it was necessary to
introduce an $SU(2)$ gauge field.  With fuller
understanding of the process of non-Abelian bosonization from
which they arise, and of exact one-dimensional models where they
are elementary excitations, it can be demonstrated that spinons
can be described as semions, and that their algebraic character and statistics do
not change all the way from the noninteracting electron case to
the Mott insulator. The spinon propagator remains asymptotically
$(x-v_{s}t)^{-1/2}$, and spinon amplitudes are SO(2) matrices with
real determinant. Thus in a Hartree-Fock factorization of the
exchange term (1) it is not consistent to allow complex
self-energies for spinons to arise. The self-energy 
can contain
an anomalous (pairing) term, but this term cannot be complex: the
self-energy is a real, orthogonal matrix, not a Hermitian one.
$$
\sum_\sim = \pmatrix{\sum & \Delta\cr 
\Delta &\sum\cr} \eqno(7)
$$
where $\Delta$ like $\sum$, is real. [Another way of seeing this
is that the two possible zero-momentum, zero-spin pairings of
spinons are 
$<s^+_{k\sigma}\ s_{k\sigma}>$ 
and 
$<s^+_{k\sigma}\ s^+_{-k-\sigma}>$, 
but 
$s_{k\sigma}$ 
and 
$s^+_{-k-\sigma}$ 
are equivalent under a $U(1)$ gauge group of Abelian rotations, which
allows only one extra degree of freedom.]

Thus the spinons can have 
a conventional self-energy $(\gamma_k)$ and a single real anomalous 
one $\Delta^d_k$, but not a
third, since no such degree of freedom exists. The existence of an
anomalous self-energy generated from the exchange term does not,
therefore, in any sense imply superconductivity, since a spinon
gap has no phase.

We can, thus, envisage two types of RVB, non-superconducting
states: one in which the spinons form a Fermi surface, with
$\Delta=0$; and one in which the spinons would have nodal points
with Dirac-like excitations about the nodes. The latter would form
below some critical temperature of the order of $J$ given by a
BCS-like equation.  Neither is stable at low $T$ because the
umklapp terms force a magnetic order on the half-filled case; but
they do have a definable order parameter in the sense that the
Fermi surface,  or
Dirac points, are an order parameter\cite{10}.

Now let us move away from the half-filled case and dope the
system. Let us assume that spin-charge separation remains
valid---for example, we can use the Zou transformation, suitably
modified to work with spinon semions and not Fermions, and from
it derive a superexchange interaction via the $t-J$
transformation. It is important to realize that the 
$\rm\underline {exchange \ interaction \ involves \ only \
spinons}$, so that any energy derived
from it still does not directly effect superconductivity; it
causes an energy gap, for spin excitations, which is intrinsically
real. (For a practical illustration of this possibility, see the
analysis in my interlayer theory of the spin gap, which is
applicable in modified form here.)\cite{11}

We shall make an arbitrary but reasonable (and I suspect
provable) assumption about charge-spin separation using spinons.
This is that as the system is doped away from half-filling, the
Fermi surface for spinons as defined by the self-energy $\sum_k$
continues to enclose a number of momentum states equal to twice the
number of particles. There are several ways to justify this.
First, the number of spin degrees of freedom = the number of 
spinons = twice the number of particles,  and independent spinons
comprise the interior of the Fermi surface only. Second, when
eventually the charge attaches itself to the spinons and makes 
quasiparticle degrees of freedom, it will be the spinons, with
their fixed statistics, which determine the Fermi surface; charge
degrees of freedom in the Luttinger liquid as in anyon theory or
the dual theories of Fisher et al have no fixed statistics.

In the Zou derivation of the $t-J$ Hamiltonian, or in fact in any
reasonable treatment of the kinetic energy term
$$T=P_G \sum_{k,\sigma} \epsilon_k\ n_{k\sigma}\ P_G$$
the $n_{k\sigma}$ may be factorized into a product of bilinear 
operators in charge and spin variables; in Zou and Baskaran's
approach as quoted in (1), (after projecting out double
occupancy):
$${\cal H} =t\sum_{<ij>\pi} e^+_i\ e_j\ s^+_{i\sigma}\
s_{j\sigma} 
+J\ \sum_{<ij>} (s^+_{i\uparrow}\ 
s_{j\downarrow}-s^+_{i\downarrow}\ s^+_{j\uparrow})\times
(cc)\eqno(10)$$

As we learn in the tomographic bosonization approach, $e$, the
``holon'' operator, is not an elementary excitation with any
conventional statistics, in the same way that $s_{i\sigma}$
is, and its dynamics need to be further analyzed. Nonetheless
$t<e^+_i\ e_j>$ must be proportional to the mean kinetic energy of
hole motion, proportional roughly to $x$, the number of holes.
This will provide an addition to the diagonal self-energy which
at half-filling is solely due
to exchange. 

The charge-correlation sum rule controlling the magnitude of the
``Drude'' peak 
is empirically, in the cuprates, and theoretically, in
1D, roughly linearly proportional to $x$; we believe the same is
to be expected for  this term in the spinon kinetic energy. Thus the spinon's
velocity, i.e., the inverse of the density of states, given by the
diagonalized self-energy, will grow proportionally to $J +
{\rm const} \times \ tx$, while $J$ itself will become smaller as $x$ 
increases, because the fraction of nearest neighbor sites which
are both occupied decreases as $(1-x)^2$. 

We can schematize these conclusions in a phase diagram of
temperature vs doping percentage $x$ (Fig 1).\cite{12}  
\   The lines are not
true phase boundaries but crossovers, since because of local  gauge
degrees of freedom actual broken symmetry can only occur at
absolute zero. We place the hypothesized transition for  $x=0$ at
$\sim J$, leaving out  of account the ordered antiferromagnetic
state, and recognize that here the ``extended $s$'' and ``d-wave''
gaps will appear together. But as the spinon velocity rises
proportionally to $x$, we will have an increasing regime where the
spinon Fermi surface is established while at a lower temperature
will appear the $d$-wave gap $\Delta^d_k$. This lower temperature
scale will be
$$T^*=J_{eff}(x)\ e^{ 
                     {  -{t_{eff}(x)\over J_{eff}(x)}  }
                    }\eqno(11)$$
where $J_{eff}(x)$ will decrease as $(1-x)^2$ and
$t_{eff}(x)$ will increase with $x$. This is presumably the
energy scale ``$T^*$'' seen in ARPES and other measurements in
the underdoped regime. The resulting smearing and gapping of the
Fermi surface is observed in the computations of Puttika et
al\cite{9}
using series extrapolations. 

Thus we have a basic phase diagram for the two-dimensional
Hubbard model which is divided into two regions, both of which
are charge-separated non-Fermi liquids. (For the two-dimensional
case, I have shown that the true Fermi liquid without separate
spinons occurs only for zero coupling or density.)
In the cuprates, as charge-spin separation becomes weaker we will
begin to get coherent interlayer coupling and a true Fermi liquid
will set in somewhere in the overdoped region. 

In the Fig (1) phase diagram, the NFL (Luttinger liquid) regime
extends over an increasingly wide region as $x$ increases.

Fig (1) is the phase diagram of the isolated plane. What I do not
presume to speculate on is whether or not this actually contains
a region of true superconductivity. The arguments of Lee Nagaosa
and Wen at
least make it plausible that this may be the case. If so, the
region where superconductivity may appear is the crosshatched area
of Fig.\ (1). As observed both by Nagaosa and Lee, and by the
``nodon'' arguments of Nayak et al, Tc for the superconductors is
very seriously limited by the low stiffness (high mass) which is
implied for gap fluctuations in the small--$x$ regime, by
straightforward sum rule arguments. Nonetheless, it appears that
superconductivity may be possible in the isolated $CuO_2$ plane,
without interlayer enhancement; several compounds seem to support
this possibility experimentally---highly doped BISCO
``$2\,2\,0\,1$'', and Tl
$2\,2\,0\,1$; in particular. (Not $NdCCO$, which is apparently an
$s$-wave superconductor.)

Optimally doped compounds, however, seem in general to develop
directly from the NFL region without an intervening $d$-wave RVB
phase (i.e., an intervening pseudogap).  However, invariably---for
the truly high Tc cases---the superconductivity occurs in the
$d$-wave state favored by RVB arguments. Also, in the
superconducting state, as opposed to the spin gap state, the nodes
experimentally are mathematically sharp, in contradiction to the
expectations of theories relying exclusively on interlayer
tunneling as expressed, for instance, in Ref.\ (7). This fact
seems to require that some component of the attractive
interaction have $d$-wave symmetry, i.e, that it be of long range
in $k$-space. Our working concept that the intralayer interaction 
only involves spinons would be sufficient, if at the optimum this
intralayer interaction is above or near its natural $T^*_c$. But
as yet it seems not clear what the admixture of interactions 
responsible for the higher Tc may be. 

It is, however, possible to make the following statement:
attractive interactions between opposite-spin electrons such as
are necessary to cause a singlet superconductor are
invariably the result of frustrated kinetic energy; and if the
mechanism is to be purely electronic (as it surely is in the
cuprates) it must be frustrated electronic kinetic energy. In a
true Landau (quasiparticle) Fermi liquid, the Fermi surface is
by definition determined in such a way a to minimize the
electronic kinetic energy as renormalized by interactions: there
is no frustrated kinetic energy. Thus there can be no attractive
interactions for opposite spins. Superexchange as in the $t-J$ model, as well as
interlayer tunneling interactions, requires frustration: the
elementary excitations must not be true quasiparticles, but
projected entities of different algebraic and possibly topological
character. We may make a small table of possible attractive
interactions:
\vskip.4truein

\begin{tabular}{|l||l|l|}
\hline
            & {\bf Intralayer}    &  {\bf Interlayer}      \\  \hline\hline
{\bf Exchange}    & Superexchange & IL Superexchange       \\  
                  &               & (Millis-Monien)        \\  \hline
{\bf Josephson}   & Pair Hopping  & IL Pair Hopping        \\  
                  & \qquad\qquad??            &                        \\
            &(charge stiffness)   &(PWA-Hsu-Wheatley)      \\  \hline
\end{tabular}

\vskip.4truein
Note that pair hopping within the plane (coherent charge motion) is not frustrated
in
the NFL state, only in the RVB state: hence we can apparently get
superconductivity in one layer if we go first into the RVB state.
But any 2D NFL frustrates interlayer hopping, hence
the IL mechanisms work from the NFL state.

One final remark needs to be made. The antiferromagnetic
insulator gains extra energy from umklapp processes, hence is a
downward cusp in energy {\it vs}  
$x$ at $x=0$. (As can be seen by the indeterminacy of
$\mu=({dE\over dx})^{-1})$. The metallic state has screening of
the long-range Coulomb interaction, which means that low density
$x$ has negative curvature ${\partial^2F\over\partial x^2}<0$. Hence
small $x$ is not a stable homogeneous thermodynamic state. On
some, presumably microscopic but occasionally macroscopic, scale
it will phase separate. Yoshimori's XPS measurements clearly
illustrate that this is happening in $(LaSr)_2 CuO_4$. This phase
separation which causes ``stripes'' is however clearly over by the optimally doped
concentration in every case. Hence it is an interesting
phenomenon, but somewhat irrelevant to our major concerns. 
It seems likely to inhibit superconductivity by trapping the
charge fluctuations, except insofar as one obtains
quasimacroscopic superconducting droplets of near optimal
concentration.

\vskip.3truein
It is not possible for me to acknowledge all the colleagues with
whom I have had useful discussions dring the past 12 years. A
list must include M. Norman, S. Chakravarty, G. Baskaran, S.
Strong, S. Kivelson, A. Adbo, W. Puttika, R. Laughlin, T. Hsu,
T.M. Rice, N.-P. Ong, but also many others. I also most
acknowledge the Aspen Center for Physics.

\vfill\eject

\end{document}